\documentclass[11pt]{article}

\usepackage{INTERSPEECH2019}
\usepackage{paralist}

\usepackage{mathtools}% http://ctan.org/pkg/mathtools
\usepackage{amsfonts}
\usepackage{amssymb,amsmath}
\usepackage{amsthm}
\usepackage{latexsym}
\usepackage{epsfig}
\usepackage{ifthen}
\usepackage{cancel}
\usepackage{esint}
\usepackage{mathptmx}
\usepackage{amsbsy}
\usepackage{bbm}
\usepackage{rotating}
\usepackage{caption}
\usepackage{multirow}

\newcommand{\beq}{\begin{equation}}
\newcommand{\deq}{\end{equation}}

\newcommand{\baq}{\begin{eqnarray}}
\newcommand{\daq}{\end{eqnarray}}

\newcommand{\baqm}{\begin{eqnarray*}}
\newcommand{\daqm}{\end{eqnarray*}}

\usepackage{tipa}
\usepackage{tipx}
\usepackage{enumitem}

\usepackage{wasysym}
\usepackage{mathptmx}
\usepackage{hyperref}
\usepackage{bm}

\usepackage{helvet}
\usepackage{courier}
\usepackage{type1cm}         

\usepackage{makeidx}         % allows index generation
\usepackage{graphicx}        % standard LaTeX graphics tool
\usepackage{multicol}        % used for the two-column index
\usepackage[bottom]{footmisc}% places footnotes at page bottom
\usepackage{chapterbib}      % Allows for local referecing of chapters

\usepackage{xcolor}          % Provides \textcolod{blue}{sample text} and
\definecolor{BeigiBlue}{RGB}{0,0,128}
\definecolor{white}{rgb}{1,1,1}
\definecolor{color-202124}{rgb}{0.13,0.13,0.14}

\ninept

\twocolumn
\title{A Transaction Represented with Weighted Finite-State Transducers\ \\ \ \\
  \small{\href{https://www.recotechnologies.com}{Recognition Technologies, Inc.}}\\
  \small{Technical Report: \href{https://www.recognitiontechnologies.com/~beigi/ps/RTI-20230131-01.pdf}{RTI-20230131-01}}\\
  \small{\href{http://dx.doi.org/10.13140/RG.2.2.35825.15205}{DOI: 10.13140/RG.2.2.35825.15205}}
}
\name{J. Nathaniel Holmes$^1$, Homayoon Beigi$^2$}
\address{
  $^1$Dept. of Appl. Phys. and Appl. Math., Columbia University\\
  $^2$Recognition Technologies, Inc. and Columbia University}
\email{$^1$jnh2139@columbia.edu, $^2$beigi@recotechnologies.com}

\begin{document}

%\definecolor{white}{rgb}{1,1,1}
\definecolor{color-202124}{rgb}{0.13,0.13,0.14}

\maketitle

\begin{abstract}
Not all contracts are good, but all good contracts can be expressed as
a finite-state transition system (``State-Transition
Contracts''). Contracts that can be represented as State-Transition
Contracts discretize fat-tailed risk to foreseeable, managed risk,
define the boundary of relevant events governed by the relationship,
and eliminate the potential of inconsistent contractual
provisions. Additionally, State-Transition Contracts reap the
substantial benefit of being able to be analyzed under the rules
governing the science of the theory of computation. Simple
State-Transition Contracts can be represented as discrete finite
automata; more complicated State-Transition Contracts, such as those
that have downstream effects on other agreements or complicated
pathways of performance, benefit from representation as weighted
finite-state transducers, with weights assigned as costs, penalties,
or probabilities of transitions. This research paper (the ``Research''
or ``Paper") presents a complex legal transaction represented as
weighted finite-state transducers. Furthermore, we show that the
mathematics/algorithms permitted by the algebraic structure of
weighted finite-state transducers provides actionable, legal insight
into the transaction.\\
\end{abstract}

\noindent\textbf{Index Terms}: Weighted Finite State Transducer, Contract, OpenFST, Tropical Semiring, Determinize, and Shortest Distance

\section{Problem Description and State of the Art}
\subsection{Contracts as Automata}
A contract is an agreement between two or more parties consisting of a
valid offer and acceptance, meeting of the minds, and
consideration. This is the definition of a contract that is recited by
students in American law schools and attorneys in American
courts. This definition is correct, but it does not describe the
``plumbing'' that makes contracts the vehicle through which we engage
in commerce.\\

Here is a better description: a contract is a relationship between two
or more parties composed of a set of reciprocal obligations and a set
of consequences for failure to perform such obligations. In most
contracts, there are numerous factual pathways to each element in the
sets of obligations and consequences, each likely carrying different
risks and/or costs to each party.\\

This description assumes the legal existence of a contract and unlike
the formal definition of a contract, describes its actual structure,
which is a defined set of mutually exclusive states and a set of
events triggering transitions between states, often defined
via-negativa. This structure can be defined with mathematical
formalism as a finite-state machine.\\

\subsection{State of the Art}
Flood and Goodenough were the first to represent contracts as
finite-state machines. They showed that a simple loan agreement can be
represented as a deterministic finite automaton (``DFA'')~\cite{r-m:flood-2002},
which is defined as a 5-tuple, (Q, ${\Sigma}$, ${\delta}$, qo, F),
consisting of the following:

\begin{enumerate}
  \item{a finite set of states, denoted Q}
  \item{a finite set of input symbols (events) called the  alphabet (${\Sigma}$)}
  \item{a transition function (${\delta}$ : Q ${\times}$ ${\Sigma}$ ${\mapsto}$ Q)}
  \item{a start state (qo ${\in}$ Q)}
  \item{a set of end states (F ${\subseteq}$ Q)~\cite{r-m:spiser-2006}}
\end{enumerate}

Utilizing the finite automaton framework, they took the provisions of
the contract, embodied in natural language ``legalese'', and reduced
them into concise labels that were placed into categories of either
events or states. Then, they gave the labels IDs and per the terms of
the contract, created a transition function that described the entire
lifecycle of the contract by showing how all the events triggered
transitions to different possible states. They illustrated the
transition function in three ways: with directed causal graphs, in
tables (with three columns - initial state, event, and resulting
state), and in a matrix form on a spreadsheet~\cite{r-m:flood-2002}.\\

Flood and Goodenough’s representation of a contract as a DFA is
significant. Computable contracts can be examined with computational
machinery, such as being visualized, tested for computational
complexity, and tested for various drafting metrics like brevity,
legal completeness, and ambiguity of
provisions~\cite{r-m:sergot-1986}. Moreover, drafting contracts as
finite-state machines eliminates risk uncontemplated by the contract,
as machines cannot invent information.\\

\subsection{Problem Description}
Flood and Goodenough suggested that future work should attempt to
represent a contract as finite-state
transducers~\cite{r-m:flood-2002}, and that is the topic of this
Research. More specifically, in this Paper, we represent a transaction
with weighted finite-state transducers (``WFST'') (\ref{def:wfst}), which have proven
very useful in other domains (e.g. speech recognition), by
transforming the natural language of the contract into states and
transitions, coding those states and transitions into a computer, and
performing computational operations~\cite{r-m:mohri-2002}. Moreover,
as noted by Flood and Goodenough, finite-state transducers expand on
DFAs by allowing transitions to emit events, which are needed to
accurately represent complex transactions (examples of transition
emitting events include a cross-default clause, which is a
\textcolor{color-202124}{provision in a loan that puts a borrower in
  default if the borrower defaults on another obligation, or a
  contract incorporating another contract or
  statute)~\cite{r-m:flood-2002}. Weights expand on the DFA
  case, by allowing the assignment of costs or penalties}
\textcolor{color-202124}{to} \textcolor{color-202124}{transitions and}
\textcolor{color-202124}{states} \textcolor{color-202124}{(e.g. the
  cost associated with a strategic default).}\\

\section{Methods and Formulation}
Similar to Flood and Goodenough’s work, we reduce the natural language
of a contract to concise labels with IDs. However, unlike the previous
research, the subject contract is longer and more complex.  Moreover, we represent the contract as WFSTs, to allow transitions to emit events and have weights in the form of costs
associated with various transitions.\\

We utilize the construct known as a semiring (\ref{def:semiring}) from abstract algebra as
the foundation of representing the transaction as a WFST. Semirings
are useful because they permit automata over a broad class of weight
sets and algebraic operations~\cite{r-m:droste-2009}. ~There are a variety of
examples of semirings. We utilize a Tropical Semiring (\ref{def:trop}) for 
reasons discussed in section 3.\\

Lastly, having reduced the contract to a WFST, we perform the
algorithms of determinization and shortest distance.\\

\subsection{Mathematical Definitions of Semiring and WFST}
\newtheorem{definition}{Definition}
\begin{definition}[Semiring]
\label{def:semiring}
A system $(S, \oplus, \otimes, \overline{0}, \overline{1})$ is a
semiring if $(S, \oplus, \overline{0})$ is a commutative monoid with
identity element $\overline{0}$; $(S, \otimes, \overline{1})$ is a
monoid with identity element $\overline{1}$; $\otimes$ distributes
over $\oplus$; and $\overline{0}$ is an annihilator for $\otimes:
\forall a \in S, a \otimes \overline{0} = \overline{0} \otimes a =w
\overline{0}$~\cite{r-m:mohri-2009,r-m:salomaa-1978}.
\end{definition}

Thus, a semiring is similar to an algebraic ring except that the additive operation may
not have an inverse~\cite{r-m:mohri-2002}. 

\begin{definition}[Tropical Semiring]
\label{def:trop}The tropical semiring has the following structure: $(R + \cup
\{\infty\}, min, +, \infty, 0)$~\cite{r-m:mohri-2009}.
\end{definition}

Weighted transducers are finite-state transducers where each
transition has a weight in addition to the input and output
labels~\cite{r-m:mohri-2009}. The weights are elements of a semiring
$(S, \oplus, \otimes, \overline{0}, \overline{1})$~\cite{r-m:mohri-2009}.
The weight of a path is computed with the $\otimes$-operation by
$\otimes$-multiplying the weights of the transitions along that
path~\cite{r-m:mohri-2009}. The $\oplus$-operation computes the weight
of a pair of input and output strings $(x, y)$ by ${\oplus}$-summing
the weights of the paths labeled with $(x, y)$~\cite{r-m:mohri-2009}. The 
formal definition follows.

\begin{definition}[Weighted Finite State Transducer]
\label{def:wfst}
A weighted transducer T over a semiring $(S, \oplus,
\otimes, \overline{0}, \overline{1})$ is an $8$-tuple $T =(\Sigma,
\Delta, Q, I, F, E, \lambda, \rho)$ where $\Sigma$ is a finite input
alphabet, $\Delta$ a finite output alphabet, $Q$ is a finite set of
states, $I \subseteq Q$ the set of initial states, $F \subseteq Q$ the
set of final states, $E$ a finite multiset of transitions, which are
elements of $Q \times$ $(\Sigma \cup \{\epsilon\}) \times (\Delta \cup
\{\epsilon\}) \times S \times Q, \lambda : I \mapsto S$ an initial
weight function, and $\rho : F \mapsto S$ a final weight function
mapping $F \mapsto S$~\cite{r-m:mohri-2002}.
\end{definition}

\subsection{Algorithms}
There are numerous algorithms permitted by the WFST structure,
however, not all likely have real application for legal analysis. Here
we present two algorithms that provide significant and immediate legal
meaning to the contract - determination and the shortest distance
algorithm.\\

A weighted transducer is deterministic if and only if each of its
states has at most one transition with any given input
sequence~\cite{r-m:mohri-2002}. In other words, if a weighted
transducer is non-deterministic, given some input the exact state to
which the machine moves (and/or the weighted path it takes) cannot be
``determined''; on the contrary, a deterministic automaton has at most
one path matching some given input.\\

Determinization takes a specific input sequence and determines its
combined weight for all its paths~\cite{r-m:mohri-2002}. This weight
calculation depends on the semiring used; here the weights are
calculated via ${\otimes}$-multiplication, which is addition in a
tropical semiring. Determinization with a tropical semiring renders
the minimum weight path as the weight associated for the given input
in the determinized automaton. For example, if in automaton A there
are two paths for input string ``xy'' with weights \{$2+3 = 5$, $5+2 =
7$\}, then the determinized equivalent weighted automaton B will
associate weight 5 with input string ``xy''~\cite{r-m:mohri-2002}. The
pseudocode of the weighted determination algorithm can be found in the
literature~\cite{r-m:mohri-2002}.\\

The shortest distance algorithm finds the shortest distance from one
state to another. These distances can be a fixed state to all other
states, such as the shortest distance between the starting state and
all other states, or the shortest distance between any states, such as
a given state and a final state. The shortest distance is determined
by minimizing the weights along the paths (${\oplus}$-sum
operation). Weights are calculated by ${\otimes}$-multiplying the
weights along the path~\cite{r-m:mohri-2004}. The pseudocode for the
shortest distance algorithm can be found in the
literature~\cite{r-m:openfst}.\\

We leave legal applications of these algorithms to Section 3.\\

\subsection{Software}
The algorithms are computed using OpenFST, which is a free library
used for manipulating WFSTs~\cite{r-m:flood-2002}. 

\section{Data, Experiments, and Results}
\label{sec:data}
\subsection{Data - Contract} 
\label{sec:data_contract}
The raw ``data'' is a contract (Section~\ref{sec:appendix}). We will 
briefly describe the
substance of the contract because it is helpful in recognizing the
applications of the algorithms that follow.\\

The contract is a manufacturing agreement between a buyer and
manufacturer of widgets.\footnote{\textsuperscript{}{This contract is based on a
publicly available contract published on SEC.gov. }} Once the parties agree to terms and the contract 
is executed, the buyer will remit a fifty percent (50\%) up-front payment to 
the manufacturer. Upon receipt of the payment, the manufacturer makes 
arrangements to produce the product before the deadline set out in the contract. 
If the manufacturer misses this deadline, the contract includes an option 
for a six week grace period for manufacturing at the option of the
buyer. If there are further delays beyond the six week point or the
optional grace period is not granted, the manufacturer is in
breach. If the product is manufactured timely, the buyer remits the
remaining half of the balance owed and the manufacturer will ship the
product to the buyer. Buyer will inspect the product upon arrival. If
the product is satisfactory, the deal is complete. If the product is
defective or nonconforming, the buyer has the option to grant the
manufacturer six weeks to cure the defects. In the event of the
manufacturer’s breach at any point in the transaction, the contract
states that the manufacturer must return all money paid by the buyer.\\

The events triggering breach are listed in the chart below. For the
purposes of assigning these breach scenarios as states, we lump them
all into one state, ``litigation.'' We assume that breach will lead to
the non-breaching party having to pursue informal or formal dispute
resolution to secure redress, despite the fact that the contract even
provides for a full refund to the buyer in the event of the
manufacturer’s breach.\\

Like Flood and Goodenough’s research, we set out the states and
transitions in table format:\\

Similar to how contracts are usually drafted by one party’s counsel
and in the favor of that party, the WFST structure analyzes the
contract from only the buyer’s perspective. As such, the costs
(weights) associated with the states and transitions are the buyer’s,
not the manufacturer’s costs.\\

We make two assumptions in the states/transitions. First, we assume that
the buyer will always grant the contractually permitted optional time
extensions in the event of delays or nonconformities. Second, we assume
that neither the buyer nor manufacturer, based on the industry
standards and their prior dealings with one another, will ever provide
opportunities to cure a breach despite being allowed by the
contract. We make note of this to stay consistent with the meaning of
``input'' and ``output'' in a finite-state machine. In other words,
some input leading to specific output is natural (for example,
{\textless}notice of timely completion : buyer remits final \$15,000
payment{\textgreater} is directly provided for in the contract)
whereas other input-output constructs may seem contrived without
explanation (for example {\textless}breach : non-breaching party
waives option to allow breaching party an opportunity to cure
breach{\textgreater}).\\

Additionally, note that twenty-seven sections of the contract are
absent from all states except via the ``litigation'' state (see \ref{sec:tables}). This
highlights a utility of representing a contract in a WFST structure -
distilling the contract into distinct parts, the substance of the deal
and the terms in the event the deal fails. This construction makes
errors more apparent to the drafter and the contract more readable to
the parties.\\

\subsection{Data - Weights}
As reflected in the Contract Transitions table, we expand on the work
of Flood and Goodenough by associating weights with the
transitions. Before discussing the weights further, note that the
purpose of this Research is to further the research of representing
contracts as finite-state machines by representing a complex
transaction as a WFST, not to examine derivation of weights, which is
typically performed with learning algorithms applied to large
datasets~\cite{r-m:mohri-2004-1}. For that reason, we did not perform
an analysis of deriving the weights and merely assigned
weights. However, this does not mean that the weights are inaccurate;
the weights are commercially reasonable
estimations of the buyer’s costs associated with transitions from
state to state.\\

We assume that the buyer engages the manufacturer to produce a single
product for a total price of $\$30,000$. We assume that every month an
order is late, the buyer loses $\$10,000$ in
profits. We assume that a breach of contract at any point in the course
of performance of the contract costs the buyer $\$30,000$ - reflecting
the lost profits during the time it will take the buyer to contract
with another manufacturer and have the product produced (this cost
does not include payment to the new manufacturer), which we estimate is
three months.\\

The transition from state 0-1 (a:b) has a weight of $\$15,000$
representing the $\$15,000$ down payment from buyer to manufacturer. The
transitions from states 1-3 (e:f) and 4-3 (e:f) have weights of $\$15,000$
representing the buyer remitting full payment (the remaining $\$15,000$)
to the manufacturer upon manufacturer's notice of timely completion of
the product. The transition from 1-4 (g:h) has a weight of $\$15,000$
representing the $\$10,000$ buyer loses per month the product is late
over the six week grace period the buyer grants to the manufacturer to
complete the product. Note we do not assume that the buyer is losing
$\$10,000$ every four weeks during the manufacturing period - only when
manufacturing is delayed beyond the due date listed in the delivery
schedule (Exhibit 3 to the contract). The transitions from 3-5 (i:j)
and 6-5 (i:j) have weights of \$0 representing no costs to the buyer in
inspecting and accepting the product and notifying the manufacturer of
its acceptance. The transition from 3-6 (k:l) has a weight of $\$15,000$
reflecting the $\$10,000$ buyer loses per month the product is late over
the six week extension period the buyer grants to the manufacturer to
cure defects in the product. The transitions from 0-2, 1-2, 3-2, 4-2,
and 6-2 (all c:d) have a weight of $\$30,000$ representing the occurrence
of a breach event and its $\$30,000$ cost.\\

We did not assign a weight to the litigation state due to the large
difference in costs of litigating different breaches. It should be
noted though that the mathematical structure of WFSTs permit assigning
weights to states in addition to transitions.\\

\subsection{Data - Drawing of the Contract as a WFST} 
\begin{figure*}[t]
  \centering
  \includegraphics[width=\linewidth]{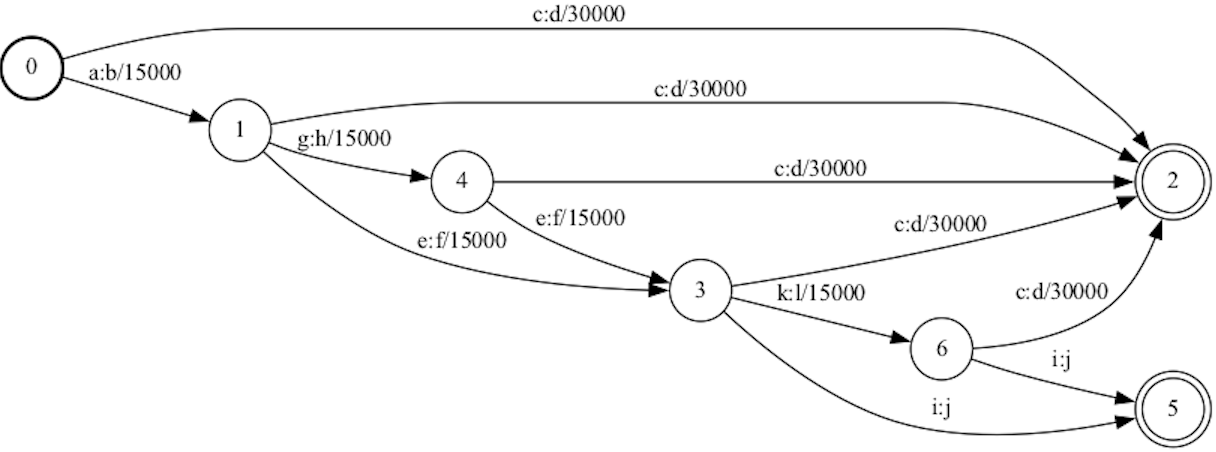}
  \caption{ Our Contract (\ref{sec:appendix}) represented as a WFST}
  \label{fig:fst}
\end{figure*}

Figure \ref{fig:fst} is the drawing of the contract represented as a WFST. Note that
the bold circle surrounding state 0 indicates that it is the starting
state and the double circles around states 2 and 5 indicate they are
final states.

\subsection{Experiments and Results} 
The determinization and shortest distance algorithms provided the most
immediate and useful analytics for the contract.\\

Determinization eliminates the possibility that the same input
(events) could give rise to multiple navigations through states. In
the context of a contract, determinization is critical; a
non-deterministic contract could have multiple, perhaps conflicting
provisions triggered by the same set of facts. This could lead to
conflict, inefficiency, and litigation.\\

Here is a simple example, without weights, for illustration. Suppose
we have a contract with the following two provisions. The first
provision requires payment within ten (10) days of receipt of some
goods. The second provision states that if the goods are
nonconforming, the buyer must provide seller notice and then there
will be an inspection period. If the inspection period lasts longer
than ten (10) days, the contract is inconsistent as it requires
payment but has not provided an adequate inspection period. Let’s
translate this scenario to states and transitions: state 1 represents
``goods received'', state 2 represents ``payment due'', state 3
represents ``nonconforming goods'', and transition A represents the
inspection period lapsing past ten days. State 1 emits transition A
into both states 2 and 3. In other words, with input of A, it cannot
be determined what state the contract is in. Determinization would
catch this inconsistency.\\

When we ran the determinization algorithm on our contract, we confirmed
that it is deterministic.\footnote{\textsuperscript{}We could have
also used the command \$fstinfo {[}filename{]}.} However, if the
contract had not been deterministic, we could have used the output to
spot the redundant paths and redrafted the contract.\\

Running the determinization algorithm on a much larger transaction
than the one presented here, such as a merger and acquisition has
great legal utility. In a M\&A, there are many parties - a buyer,
seller, attorneys on both sides, investment bankers, and other
specialists; the transaction also requires a lot of documents,
including employee rosters and contacts, contingent litigation,
schedules showing assets and inventory, schedules of intellectual
property, collateral agreements, true-ups, and etcetera. The many
parties contributing to drafting the terms and the number of documents
involved makes it likely that there are inconsistent provisions that
could be caught using determinization.\\

Determinization also has great application to other complex legal
documents, such as statutes and codes where ambiguities and
inconsistencies could lead to expensive litigation, commercial
inefficiency, and public policy issues.\\

Now, we move to the shortest distance algorithm. As stated above, the
shortest distance algorithm finds the cheapest routes between
states. There are two useful applications of this algorithm to the
contract.\\

\begin{figure}[t]
  \centering
  \includegraphics[width=\linewidth]{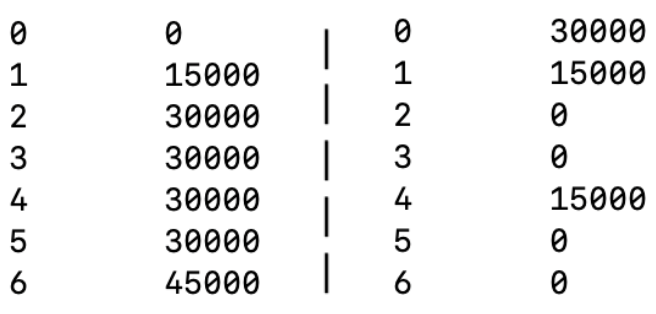}
  \caption{Outputs of the shortest distance algorithm}
  \label{fig:table}
\end{figure}

First, we computed the shortest distance from the initial state. This
translates to the cheapest cost for citizenship in each state. In
other words, what is the least a state will cost based on the
construction of the contract and weights assigned as costs for various
transitions. The output is to the left of the dashed line in figure \ref{fig:table}. We
observe that the shortest path algorithm assigns 30,000 as the
shortest path to state 5, which is correct because if everything goes
off without a hitch in the deal, the buyer ends up in state 5 (TERM)
having only spent \$30,000. We can confirm all the other output
similarly.\\

Second, we computed the shortest distance from each state to a final
state. The output is to the right of the dashed line in figure \ref{fig:table}. Like before, we
can verify the accuracy of the algorithm by confirming that the
cheapest path from each state to a final state is the same as the
output.

Both applications have important applications to contracts. The
shortest distance from the initial state tells a party the best case
cost it should expect given a factual scenario (an input). The
shortest distance from each state to a final state tells a party that
is in the middle of deal their costs to end the deal, whether that is
performance or a strategic breach. Like determinization, in small
contracts errors or choices may seem obvious, but in a large
transaction with hundreds of variables, costs, and changing
probabilities of events, it would be very useful to have this
computational horsepower available.\\

\section{Conclusion and Proposed Future Work}
Here, we provided proof of concept for representing~commercial
transactions as WFSTs. This transformation from natural language to
WFSTs permitted analysis using the tools of computation. Also, we
showed that the determination and shortest path algorithms provide
meaningful legal insight. There is more work to be done, though.\\

First, here we only used the determination and shortest distance
algorithms. This only scratches the surface of the computational
algorithms available. While we did investigate many other algorithms in
this Research, further work should be dedicated to analyzing whether
other algorithms permitted by the contract as WFSTs structure provides
valuable legal insight/application.\\

Second, here we analyzed only one type of contract - a contract for the
sale of goods. There are many different types of transactions and
legal documents, such as statutes, that would benefit from
representation as a weighted finite-state transducer. Moreover,
algorithms that provide little or no insight to the contract analyzed
here might provide insight into other transactions.\\

Third, the weights used here modeled costs associated with
transitions. We did this because weights as costs provided insight into
the financial consequences for the buyer related to each jump from
state to state within the contract. However, other weights, such as
probabilities, may be used with success and there are semiring
constructions available for such weights. Moreover, experimentation
with other semiring constructs might provide valuable legal insight.\\

Lastly, here we defined the states, transitions, and weights
manually. It would be very useful if some or all of this process could
be expedited by interfacing natural language processing (NLP) technology
with the contract as WFSTs construct. For example, NLP, perhaps a text 
summarization model specifically trained on text data from legal documents 
or laws, could extract the meaning and sentiment behind the legalese and
reduce the text into concise natural language equivalents with labels, as done
here. Then states, transitions, and weights could be assigned manually as a
first step, greatly reducing the time required for setting up a contract in a WFST 
structure.\\

%\newpage

\bibliographystyle{IEEEtran}
\bibliography{ms}

\onecolumn
\renewcommand{\arraystretch}{1.3}
\section{Appendix A: Tables}
\label{sec:tables}
\begin{table}[hbt]
\begin{center}
\caption{{\it Breach Events}}
\label{tab:breach events}
\begin{tabular}{|c|c|}
\hline
{\bf Breach Event} & {\bf Contract Section}  \\
\hline
\hline
Products insufficient quality and quantity & $1$ \\
\hline
Products not in compliance with standards and warranties & $1$(a) \\
\hline
Manufacturer does not provide parts, labor, or materials & $1$(b) \\
\hline
Manufacturer does not make its facility and product available for inspection & $1$(c) \\
\hline
Manufacturer does not provide QC or product information upon request & $1$(d) \\
\hline
Manufacturer utilizes unauthorized subcontractors and suppliers & $1$(e) \\
\hline
Manufacturer does not provide batch and lot codes & $1$(f) \\
\hline
Manufacturer does not provide certificate of analysis & $1$(g) \\
\hline
Manufacturer does not provide date of manufacturer on products & $1$(h) \\
\hline
Manufacturer does not maintain manufacturing certifications or GMPs & $2$(a) \\
\hline
Manufacturer does not maintain emergency action plan & $2$(b) \\
\hline
Manufacturer does not assist in product enhancement and product development & $2$(c) \\
\hline
Manufacturer does not provide management supports & $2$(d) \\
\hline
Manufacturer does not provide assistance with product development in developing markets & $2$(e) \\
\hline
Manufacturer does not make its facility available for inspection & $2$(f) \\
\hline
Manufacturer does not comply with price increase/price decease procedures & $3$(a), (b)\\
\hline
Manufacturer does not remit down payment or final payment & $4$ \\
\hline
Manufacturer does not meet manufacturing requirements (e.g. compliance manufacturing laws) & $6$(a)\\
\hline
Product does not comply with laws in target market & $6$(b)\\
\hline
Product does not comply with labeling requirements & $6$(c)\\
\hline
Breach of delivery terms & $7$\\
\hline
Delay of delivery of product, including after six week extension period & $8$\\
\hline
Manufacturer does not provide and maintain an inspection procedure and quality assurance program & $10$\\
\hline
Buyer or Manufacturer breach of confidentiality program & $11$\\
\hline
Buyer IP infringement & $12$\\
\hline
Seller IP infringement & $12$\\
\hline
Manufacturer failure to notify of inspection event & $13$\\
\hline
Manufacturer failure to notify of return or recall & $14$\\
\hline
Manufacturer failure to notify of regulatory action & $15$\\
\hline
Buyer or seller: bankruptcy, liquidation, government action (including litigation), or material breach & $16$\\
\hline
Manufacturer failure to maintain insurance & $17$\\
\hline
\end{tabular}
\end{center}
\end{table}

\vspace{-2em}

\begin{table}[ht!]
\begin{center}
\caption{{\it Natural Language Labels and State Assignments}}
\label{tab:state_assignments}
\begin{tabular}{|c|c|c|}
\hline
{\bf State} & {\bf Natural Language Label} & {\bf Contract Section} \\
\hline
\hline
$0$ & START & n/a \\
\hline
$1$ & production period has elapsed & $8$ \\
\hline
$2$ & litigation & $9, 18-37$ \\
\hline
$3$ & produce shipped & $4, 7$ \\
\hline
$4$ & six week production extension period elapses & $8$ \\
\hline
$5$ & TERM/contract complete & $n/a$ \\
\hline
$6$ & "cure period" has elapsed & $8$ \\
\hline
\end{tabular}
\end{center}
\end{table}

\vspace{-2em}

\begin{table}[ht!]
\begin{center}
\caption{{\it Transition IDs and Natural Language Labels}}
\label{tab:transitions}
\begin{tabular}{|c|c|c|c|}
\hline
{\bf Transition IDs} & {\bf Natural Language Label} & {\bf  Contract Section} & {\bf  Weights}  \\
\hline
\hline
a:b & signed contract: $\$15,000$ payment & $4,5$ & $15,000$ \\
\hline
\multirow{2}{*}{c:d} & breach event* : non-breaching party waives option & \multirow{2}{*} {*, $18$} & \multirow{2}{*} {$30,000$} \\
& to let breaching party cure breach &  & \\
\hline
e:f & notice of timely completion : buyer 
remits final 50\% payment &  $4, 8$ & $30,000$ \\
\hline
g:h & "delay" event : grant of extension & $8$ & $15,000$ \\
\hline
\multirow{2}{*}{i:j} & buyer inspects product and accepts : manufacturer
receives & \multirow{2}{*}{$8$} & \multirow{2}{*}{$0$} \\
& notice of buyer's acceptance & & \\
\hline
k:l & buyer rejects product : buyer grants manufacturer
opportunity to cure &  $10$ & $15,000$ \\
\hline
  & *see breach events table &  &  \\
\hline
\end{tabular}
\end{center}
\end{table}

\renewcommand{\arraystretch}{1}

\onecolumn
\section{Appendix B: Manufacturing Agreement}
\label{sec:appendix}
THIS MANUFACTURING AGREEMENT (the “Agreement”) dated as of the [date], by and between [insert buyer], a [insert state/country] corporation having a place of business at address (“Buyer”), and [insert seller/manufacturer], a [insert state/country (“Manufacturer”) having a place of business at [insert address] (each a “Party” and collectively the “Parties”).

\begin{enumerate}
  \item{\underline{{\bf Manufacturing}}. The detailed formulations and
    specifications for the manufacturing, producing and packaging of
    all products (the “Products”) shall be listed
    on subsequent written memorandums signed by the Parties and
    expressly referring to this Agreement (the
    “Standards”). Manufacturer agrees to produce and deliver Products
    in sufficient quantity and quality in accordance with the terms
    and provisions of this Agreement.}
    \begin{enumerate}[label=\alph*.]

      \item{\underline{{\bf Compliance with Standards and Warranties}}. Manufacturer
        shall produce the Products in accordance with the
        Standards. Manufacturer shall conduct in-process inspections,
        final inspection and perform testing as mutually agreed upon
        by the Parties to insure that all Products are manufactured in
        compliance with the Standards. Manufacturer shall not make any
        changes in the specifications or formulations without the
        prior written consent of Buyer. All Products manufactured for
        Buyer by Manufacturer shall be manufactured and delivered in
        accordance with the warranties contained in Section 6. In
        order to insure compliance with this Agreement, Manufacturer
        shall maintain a retained sample of each batch and lot of
        Products produced by Manufacturer for a period of five (5)
        years from the production date.}
        
        \item{\underline{{\bf Parts, Labor, and Materials}}. Manufacturer shall provide all parts, labor, and materials necessary to perform Manufacturer’s obligations under the terms of this Agreement. Manufacturer shall maintain, at no cost to Buyer, an inventory of raw materials used in the manufacture of the Products reasonably sufficient to meet Buyer’s Purchase Order (defined in Exhibit 1). Manufacturer shall maintain such inventory on a FIFO basis.}

        \item{\underline{{\bf Inspection Rights}}. Buyer or its representatives may review Manufacturer’s performance of the work under this Agreement including development, formulation, production and tests of the Products, the design of the manufacturing process used to produce them, and their operation. To review the work, Buyer or its representatives may visit the sites where Manufacturer and/or Manufacturer’s subcontractors and agents perform the process, or Buyer or its representatives may review any and all documentation related to Manufacturer’s performance of the work hereunder. Buyer may review such documentation at Manufacturer’s site or request Manufacturer to provide copies for review. Buyer shall visit the sites during normal business hours and shall have access to documentation with reasonable notice to Manufacturer.}
        
         \item{\underline{{\bf Quality Control and Production Information}}.
         Upon receipt of a request for information relating to formulation, sources of ingredients, suppliers, subcontractors or other information relating to the Products, Manufacturer shall provide all requested information and cooperate fully and to the extent reasonably requested with the party requesting such information.}

	\item{\underline{{\bf Subcontractors and Suppliers}}.
	Manufacturer currently utilizes certain subcontractors and suppliers in order to perform its obligations hereunder whose names are listed on a written memorandum signed by the Parties and expressly referring to this Agreement. Manufacturer shall not utilize any subcontractors or suppliers other than those listed in such written memorandum in the manufacturing process without obtaining the prior written consent of Buyer to such additional or replacement subcontractors or suppliers. Manufacturer shall not be required to obtain the consent of Buyer before changing subcontractors or suppliers involved solely in shipping portion of the manufacturing process.}
	
	\item{\underline{{\bf Batch and Lot Codes}}.
	Each Product manufactured by Manufacturer under this Agreement shall be identified by a lot number that is linked to the manufacturing Batch Number of the Product and location, time and shift of final packaging. The term “Batch Number” shall mean a number which is assigned to a single production run of a Product manufactured by Manufacturer.}
	
	 \item{\underline{{\bf Certificates of Analysis}}.
	 Manufacturer shall ensure that an appropriate certificate of analysis accompanies each shipment of Products to Buyer. If, at Buyer’s request, Products are shipped to a third-party distributor of Buyer, Manufacturer shall provide to both Buyer and the entity receiving the shipment a certificate of analysis. In either case, the certificate shall, at a minimum, provide an analysis of the Products contained in that shipment, as well as the input amounts of all components of the Products with label claims, and the results of all assays performed) and the bar coded information in the form set forth in the Standards. Buyer or any recipient of a shipment shall have the right to reject any shipment of Products if such shipment is received by Buyer or other recipient without a certificate of analysis, provided however that Manufacturer shall be given notice of any missing certificate of analysis and three (3) business days to deliver the missing certificate to Buyer or such third party before any such rejection can occur. Manufacturer is also responsible to maintain certificates of analysis from all suppliers of materials blended into the Products, and to insure that these conform to Buyer’s and Manufacturer’s agreed upon specifications for the Products.}
	 
	\item{\underline{{\bf Date of Manufacture}}.
	Each Product manufactured by Manufacturer under this Agreement shall display a date of manufacture consisting of month and year on the label, container, or product.}
\end{enumerate}

 \item{\underline{{\bf Consulting and Other Services}}. In consideration for Buyer utilizing Manufacturer’s manufacturing services, in addition to the manufacturing services described above, Manufacturer will also provide the following services to Buyer:}
    \begin{enumerate}[label=\alph*.]	

	\item{\underline{{\bf Certifications and Good Manufacturing Practices}}.
	Manufacturer shall maintain the appropriate manufacturing certifications, to be mutually agreed on between Manufacturer and Buyer.}
	
	\item{\underline{{\bf Emergency Action Plan}}.Manufacturer shall maintain an Emergency Action Plan (“EAP”) reasonably agreeable to Buyer that enables Manufacturer to respond to Buyer forecast volume requirements in the event of a business disruption. Manufacturer shall maintain offsite backup copies of all Standards, documentation, formulas, specifications, vendor listings and any other data necessary to begin manufacturing Products immediately after a business disruption and in accordance with the EAP. Manufacturer shall also enter into any necessary agreements with ingredient, raw material and packaging suppliers to ensure that the terms of the mutually agreeable EAP can be met within timeframes specified in the EAP.}
	
	\item{\underline{{\bf Product Enhancement and New Product Development}}.
	Manufacturer shall provide all reasonable assistance necessary to Buyer to enhance existing products including consulting on raw material processes and proposed formula changes as well as assisting in evaluating potential changes to ingredients or raw materials. Manufacturer shall also assist in the development of new products or the expansion of existing products to new markets including, but not limited to, regulatory consulting.
}

	\item{\underline{{\bf Management Support}}.
	Manufacturer shall provide reasonable management support to assist in the resolution of issues that may arise from time to time with respect to product questions, registrations, ingredients, disputes with governmental agencies, import or export agencies and any other entities as may be requested from time to time by Buyer.
}

	\item{\underline{{\bf Developing Markets}}.Manufacturer shall provide reasonable support to Buyer to develop new international markets including regulatory consulting, product formulation consulting, clinical study consulting and any marketing experiences Manufacturer may have.}
	
		\item{\underline{{\bf  Facility Tours - Monthly and Special}}. Manufacturer shall cooperate with Buyer to provide tours of Manufacturer’s receiving, production, packaging and laboratory facilities. On a quarterly basis, Manufacturer and Buyer will coordinate no less than three days designated for tours of the production facility by Buyer. Special tours may be arranged from time to time for special events or personnel requested by Buyer and reasonably agreed to by Manufacturer. Manufacturer shall be responsible for providing adequately trained guides for each tour, any direct costs related to the tour on Manufacturer’s premises and ensuring Manufacturer’s facilities are adequately prepared for each tour. To the extent practical, Manufacturer will attempt to schedule production in such a way that participants in the tour will see Products covered by this Agreement being produced and packaged. }
	  \end{enumerate}
	  
	   \item{\underline{{\bf Purchase Price}}. The purchase price to be paid by Buyer for each Product shall be listed on subsequent written memorandums signed by the Parties and expressly referring to this Agreement. Any change in the purchase price is subject to the following:}
	   
    \begin{enumerate}[label=\alph*.]
    
     \item{\underline{{\bf Price Increases}}. In the event that manufacturing and other Product related costs increase materially as a result of labor costs, material costs, rent, custom charges, state taxes, import or export fees, freight costs, utility rates, or other costs, Manufacturer shall provide Buyer documentation supporting such cost increases in a form reasonably satisfactory to Buyer. Upon Buyer’s reasonable satisfaction and confirmation of the increased manufacturing and other Product related costs, the increased costs shall be reflected in an increased purchase price of the Products paid by Buyer to Manufacturer on a per Product basis to be determined by Buyer and Manufacturer. In such an event the purchase price of the Products shall increase by a percentage equal to the percentage of the increase in Manufacturer’s manufacturing and other Product related costs. Any increase in purchase price shall become effective ninety (90) days after such increase is determined by Buyer and Manufacturer. Manufacturer agrees to take all customary and reasonable steps to maintain manufacturing costs at levels consistent with or below such costs as of the date of this Agreement. For purposes of this Section, manufacturing and other Product related costs shall be examined annually, with the first such examination to occur on the first business day after [insert date].}
  
   \item{\underline{{\bf Price Decreases}}. In the event manufacturing costs and other Product related costs decrease materially, Manufacturer shall inform Buyer of such decrease and negotiate with Buyer, in good faith, a reduction in the purchase price of each Product. The decreased costs shall be “passed through” to Buyer on a per Product basis to be determined by Buyer and Manufacturer, with such decrease to reflect a direct pass through of such decreased manufacturing costs. Any decrease in purchase price shall become effective ninety (90) days after such decrease is determined by Buyer and Manufacturer. For purposes of this Section, manufacturing costs shall be examined annually, with the first such examination to occur on the first business day after [insert date].}
   
    \end{enumerate}
   
    \item{\underline{{\bf Payment Terms}}. Accompanying execution of this Agreement, Buyer shall remit fifty percent (50 percent) of the Purchase Price (defined in Exhibit 2) (the “Down Payment”) to Manufacturer. Payment in full shall be due thirty (30) days after Manufacturer notifies Buyer of completion of the Products.}
    
      \item{\underline{{\bf Purchase Price}}. The purchase price (“Purchase Price”) shall be set out in Exhibit 2.}
      
       \item{\underline{{\bf Product Requirements}}. }
        \begin{enumerate}[label=\alph*.]
        
      \item{\underline{{\bf Manufacturing}}. Products manufactured by Manufacturer (i) shall be manufactured in conformity with the Standards and comply in all respects to all applicable laws of the intended marketplace, (ii) will have a shelf life equal to or in excess of the shelf life specified in the Standards, and (iii) Manufacturer, except as set forth in Section 6(b), shall not change any formulation or specification for the Products without the prior written consent of Buyer, which consent may be withheld in Buyer’s sole and absolute discretion. All Products sold hereunder shall be of merchantable quality, free from defects, fully acceptable, fit for their intended use and manufactured in conformity with the Standards and comply in all respects to all applicable laws, regulations, statutes and orders of the intended marketplace, and any intended marketplace in which (i) Buyer advised Manufacturer prior to manufacture and delivery, in writing, the Products are to be sold and in which (ii) Manufacturer participated in or reviewed the procurement of any necessary governmental registrations or approvals. Each Product shall be delivered free and clear of all liens, security interests, and/or encumbrances of any type or nature.}
      
      \item{\underline{{\bf Legal Requirements}}. Should applicable law requirements specify defect limits or other requirements that are more stringent than those, if any, contained in the Standards, the more stringent requirements shall prevail and apply and the Standards shall be automatically modified without the requirement of action by either Party. Notwithstanding the foregoing, Manufacturer shall not change any Standards as a result of the preceding sentence without the prior written consent of Buyer. In the event Manufacturer and Buyer fail to agree on any modification that Manufacturer deems required under this Section, Manufacturer shall not be obligated to manufacture any Product in accordance with any Standard that Manufacturer deems to be non-conforming, and the Parties shall negotiate in good faith to resolve the issue.}
      
       \item{\underline{{\bf Labeling}}. All packaging and labeling provided by Manufacturer for Products manufactured by Manufacturer under this Agreement shall be in conformity with the Standards and comply in all respects to all applicable laws, regulations, statutes and orders of the intended marketplace and any intended marketplace in which (i) Buyer advised Manufacturer prior to manufacture and delivery, in writing, the Products are to be sold and in which (ii) Manufacturer participated in or reviewed the procurement of any necessary governmental registrations or approvals. No Product contained in any shipment now or hereafter made to Buyer will, at the time of such shipment or delivery, be adulterated, mis-labeled or misbranded within the meaning of any applicable law, ordinance, rule or regulation, in existence at the time of shipment or delivery.}
       
       \item{\underline{{\bf Continuing Effect}}. The representations, warranties, and covenants contained herein shall be continuing representations, warranties, and covenants and shall be binding upon Manufacturer with respect to all Products that Manufacturer ships or delivers to Buyer or its designee.}
       
        \end{enumerate}
        
        \item{\underline{{\bf Delivery of Products}}. Manufacturer shall ship Products to the Buyer within one business day upon receipt of payment in full. Manufacturer shall provide Products to the Buyer in the location that is set forth in the Purchase Order. In the event Manufacturer is requested to ship Products on Buyer’s behalf, Manufacturer shall deliver the Products to the party and the final destination set forth in the Purchase Order. It is the responsibility of Manufacturer to schedule production and delivery of all Products ordered under this Agreement.}
        
         \item{\underline{{\bf Delay}}. In the event of delay, defined as the event where the Products are not delivered by the delivery date set out in Exhibit 3 (“Delay”), Manufacturer, at the express approval of Buyer, may be granted a six (6) week extension to deliver the Products to Buyer (the “Extension Period”). If Buyer grants Manufacturer the Extension Period and Manufacturer does not deliver the Products during the Extension Period, Manufacturer will refund the “Down Payment” to Buyer.}
         
         \item{\underline{{\bf Title and Risk of Loss}}. The title to and all risk of loss of the Products shall remain with Manufacturer until loaded onto the designated shipper.}
         
         \item{\underline{{\bf Acceptance and Rejections}}.  Manufacturer shall provide and maintain an inspection procedure and quality assurance program for the Products and their production processes. All inspection records maintained by Manufacturer shall be made available to Buyer, at a reasonable time, upon request. Buyer and any Master Distributor to whom Products are shipped by Manufacturer shall have fifteen (15) calendar days from the date of delivery to inspect and test all Products and may refuse to accept Products which do not conform to the Standards. If Buyer or such Master Distributor has not timely notified Manufacturer of rejection, then the Products shall be deemed to have been accepted by Buyer. The act of payment for Products shall not of itself signify acceptance. Buyer or any Master Distributor to whom Products are shipped by Manufacturer shall have the right to reject any Products delivered to Buyer or such Master Distributor which are not accompanied by or preceded by a certificate of analysis, as described in Section 1(g).  While inspecting Products, Buyer, its Master Distributors, distributors or any other representative or agent of Buyer shall store all shipped Products in clean space suitable for storage of food and protection of its contents with respect to integrity and quality, in compliance with good commercial practice, the Standards and all applicable laws, rules and regulations of the intended marketplace. If Buyer rejects Product, it may at its express approval give Manufacturer a six (6) week period to cure the defects in the Product (“Cure Period”). If Manufacturer does not complete the Product per the Purchase Order and conforming to all standards during Cure Period, Manufacturer is in breach of this Agreement. }
         
                  \item{\underline{{\bf Manufacturer's Access to Confidential Information}}. The Parties agree and acknowledge that as a result of this Agreement, each party shall receive and have access to information, including, without limitation, information regarding the Product specifications and formulations, costs of manufacture, pricing, and information regarding customers, which is proprietary to and a trade secret of the other party and which is governed by this Section, all of which shall be considered “Confidential Information.” Each party covenants and warrants to the other party that it shall not disclose or divulge Confidential Information except to the extent: (i) required by law, (ii) to protect its interests in any dispute or litigation, (iii) necessary to perform its obligations under this Agreement, or (iv) if such information becomes publicly available without breach of this Section. The Parties’ obligations under this Section shall survive any termination or expiration of this Agreement.}
                  
                   \item{\underline{{\bf Intellectual Property, Formulations, and Suppliers}}. Buyer hereby warrants that it is the owner or exclusive licensee of the formulations for the Products that are the subject of this Agreement and that it has the right to manufacture or have manufactured such Products, and Manufacturer acknowledges Buyer’s rights in the Products. Manufacturer shall not be permitted to use the formulations for the Products in any way except as necessary to perform its obligations under this Agreement. Manufacturer acknowledges Buyer’s exclusive ownership of the trademarks affixed to and any patents embodied in the Products and will do nothing at any time, during or after the term of this Agreement, which could adversely affect their validity or enforceability, including any modification or obliteration of the trademark or patent markings on the Products as sold. This Agreement shall not give Manufacturer any right to use the “[insert company name of Buyer]” or “[insert trademark name®]” name, logo, and marks, or any other trademarks of Buyer, except as specifically authorized by Buyer. Promptly following the termination of this Agreement for any reason, Manufacturer agrees to discontinue use of the “[insert company name of Buyer]” and “[insert trademark name®]” marks, and any other Buyer names and trademarks and to remove, or dispose of, as Buyer shall direct, any signs or other indicia relating to Buyer’s name and trademarks. Following termination of this Agreement, Manufacturer shall not be permitted to use the “[insert company name of Buyer]” or “[insert trademark name®]” name, logo or marks on any other Buyer name or trademark in connection with any product. Manufacturer shall not have any right to register any trademarks identical with or similar to Buyer’s trademarks. All use of Buyer’s trademarks by Manufacturer in connection with this Agreement shall be subject to Buyer’s control and shall inure to the benefit of Buyer. Buyer hereby licenses to Manufacturer during the term of this Agreement the use of the “[insert company name of Buyer]” and “[insert trademark name®]” trademarks and other intellectual property rights solely for Manufacturer’s use in the manufacture and sale of the Products to Buyer. Any and all improvements, modifications, inventions or discoveries by Manufacturer or its employees relating to the Products and formulations shall be the sole and exclusive property of Buyer. Manufacturer’s obligations under this Section shall survive any termination or expiration of this Agreement. The failure by Manufacturer to adhere to any of the terms of this Section  shall be a material breach of this Agreement.}
                   
                   \item{\underline{{\bf Inspection Events}}. Manufacturer shall immediately notify Buyer by the most expeditious means practicable, but in no event later than the next business day, if and when it is informed of an impending audit, inspection and/or onsite visit (“Inspection Event”) concerning the manufacture of any Product by Manufacturer under this Agreement by a governmental agency or any licensing unit thereof. Buyer, at its sole discretion and expense may elect to send an employee or designee to observe the Inspection Event. In the event that Manufacturer should not have prior notice of an Inspection Event, then Manufacturer shall immediately, but in no event later than the next business day after such Inspection Event, give written notice of the same to Buyer, and shall further provide to Buyer any written documentation supplied to Manufacturer on account of such Inspection Event. In the event of any action described in this Section, the Parties shall cooperate in determining the response, if any, to be made to such action.}
                   
                \item{\underline{{\bf Returns and Recalls}}. Manufacturer shall immediately provide Buyer with notification of any event or occurrence that could necessitate the need to recall or withdraw Products together with such information as may be available to Manufacturer concerning the degree to which the reasons may have application to any Products shipped to or on behalf of Buyer. In the event of such event or occurrence, Manufacturer may request the return of any such Products in the possession of Buyer or its Master Distributors. Buyer shall manage all recall decisions with respect to Products sold or shipped by it to its Master Distributors and/or customers.}
                
                 \item{\underline{{\bf Regulatory Action}}. If any government agency makes, with respect to any Product manufactured by Manufacturer for Buyer under this Agreement, (i) an inquiry, or (ii) gives notice of or makes an inspection at any party’s premises, or (iii) seizes any such Product or requests a recall, or (iv) directs any party to take or cease taking any action, the other party shall be notified immediately but in no event later than the next business day. Manufacturer will investigate the inquiry or complaint and provide Buyer with a written report within three (3) business days after the notification. Duplicates of any samples of Product taken by such agency shall be sent to the other party promptly. In the event of any action described in this Section, the Parties shall cooperate in determining, and will mutually agree upon, the response, if any, to be made to such action and each party agrees to cooperate with the other in responding to any communication or inquiry and/or attempting to resolve any such action.}
                    
                 \item{\underline{{\bf Termination}}. This Agreement may be terminated upon the occurrence of the following; if buyer or seller: (a) becomes insolvent or has a petition in bankruptcy, reorganization or similar action filed by or against it; (b) has all or a substantial portion of its capital stock or assets expropriated or attached by any government entity; (c) is dissolved or liquidated or has a petition for dissolution or liquidation filed with respect to it;(d) is subject to property attachment, court injunction, or court order materially affecting its operations under this Agreement; or (e) breaches any representation, warranty, covenant, obligation, commitment or other agreement contained in this Agreement provided, however, that, notwithstanding anything else to the contrary contained herein, in the event of a material breach by Manufacturer of its obligations under the Sections above if the breaching party delivers to the non-breaching party a written notice specifying the nature of the default and the breaching party fails to cure such default within three (3) business days following the delivery of such notice, then and only then shall the non-breaching party have the right to terminate or cancel this Agreement without further opportunity to cure.}

	 \item{\underline{{\bf Insurance}}. Manufacturer will, at Manufacturer’s expense, maintain in full force and effect, products liability insurance coverage with a policy limit of at least [insert \$amount] per occurrence and [insert \$amount] in the aggregate, consisting of at least [insert \$amount] in primary coverage and the remaining [insert \$amount] in an umbrella form for excess liability coverage. Such policy referred to in this Section shall (a) name Buyer and any master distributor or affiliated company designated by Buyer as additional insured parties thereunder (without any representation or warranty by or obligation upon Buyer) as respects distribution or sale of Manufacturer’s products, (b) provide that at least thirty (30) days prior written notice of cancellation, amendment, or lapse of coverage shall be given to Buyer by the insurer, (c) provide worldwide coverage for occurrences; and (d) provide coverage for occurrences during the term of this Agreement which will continue for such occurrences after the term of this Agreement. Manufacturer will deliver to Buyer original or duplicate policies of such insurance, or satisfactory certificates of insurance.}
	 
	 \item{\underline{{\bf Breach}}. The failure by the Parties to adhere to any of the terms of Sections 1-4, 6-8, and 10-17 above shall be a material breach of this Agreement (“Breach”). Upon occurrence of a Breach, the non-breaching party has the option, but not the obligation, to let the breaching party attempt to cure its breach. In the event the non-breaching party gives the breaching party the opportunity to cure its breach, the parties must agree in writing to terms and conditions for curing the breach. In the event of Manufacturer’s breach, including after the Manufacturer has been granted an opportunity to cure the breach but has failed, the Manufacturer must refund the buyer in full within three business days of Buyer’s request.}

	 \item{\underline{{\bf Election to Continue}}. As stated in Section 18 above, in the event of a default and the lapse of any applicable cure period, the non-defaulting party may agree to continue the Agreement rather than terminating it. By agreeing to continue the Agreement in this manner, the non-defaulting party does not waive its right to later terminate the Agreement for default based on the event of default that is the subject of the notice.}

	\item{\underline{{\bf Force Majeure}}. Neither party shall be in default nor liable to the other for any failure to perform directly caused by events beyond that party’s reasonable control, such as acts of nature, labor strikes, war, insurrections, riots, acts of governments, embargoes and unusually severe weather provided the affected party notifies the other party within ten (10) days of the occurrence. Such an event is an Excusable Delay. The party affected by an excusable delay shall take all reasonable steps to perform despite the delay. If the party is unable to perform within a reasonable period, this Agreement shall end without any further obligation of the unaffected party.}
	
	\item{\underline{{\bf Return of Materials}}. If Buyer terminates this Agreement, Manufacturer shall complete all work in process in a timely fashion and deliver the same to Buyer as provided herein against payment as provided herein. To the extent that after such work in progress has been completed, Manufacturer has inventory of raw materials and packaging materials on hand that were purchased in good faith reliance upon the rolling forecasts, then Buyer shall be liable for, and required to purchase such inventory from Manufacturer within thirty (30) days from the date that Manufacturer furnishes to Buyer a written reconciliation showing the amount of such inventory; provided that such inventory is in compliance with the Standards. With the approval of Buyer, Manufacturer may try to use all or any part of such inventory for other customers or sell all or any part of it to third parties. The Parties shall cooperate and utilize their reasonable best efforts to prepare such final reconciliations of Products and inventory and any other amounts to be provided as between them in connection with such termination. Upon payment of all amounts owed to Manufacturer, Manufacturer shall return to Buyer all materials containing the Confidential Information, documents produced in the performance of this Agreement, work-in-process, parts, tools and test equipment paid for, owned or supplied by Buyer.}
	
	\item{\underline{{\bf Amendments}}. This Agreement may only be changed or supplemented by a written amendment, signed by authorized representatives of each party.}
	
	\item{\underline{{\bf Assignment}}. Neither party may assign its rights or delegate its obligations under this Agreement without the prior written approval of the other party. Any attempted assignment or delegation without such an approval shall be void. Provided, however, that Buyer may assign this Agreement to any Affiliate of Buyer, without being released from its obligations hereunder. “Affiliate” shall mean any individual or entity that directly or indirectly controls, is controlled by, or is under common control with Buyer.}
	
	\item{\underline{{\bf Governing Law and Forum}}. This Agreement shall be governed by the laws of the State of [insert state] without regard to any provision (including conflicts of law provisions) which would require the application of the law of any state other than the State of [insert state]. All disputes arising under or in connection with this Agreement shall be determined by actions filed in the courts within the State of [insert state].}
	
	\item{\underline{{\bf Severability}}. If any provision of this Agreement is held to be illegal, invalid or unenforceable by a court of competent jurisdiction, the remaining provisions shall not be affected.}
	
	\item{\underline{{\bf Effect of Title and Headings}}. The title of the Agreement and the headings of its Sections are included for convenience, and shall not affect the meaning of the Agreement or the Section.}
	
	\item{\underline{{\bf Waiver}}. Failure of either party to insist in any strict conformance to any term herein, or in Purchase Orders issued hereunder, or failure by either party to act in the event of a breach or default shall not be construed as a consent to or waiver of that breach or default or any subsequent breach or default of the same or any other term contained herein.}
	
	\item{\underline{{\bf Indemnification by Buyer and Manufacturer}}.}
	
	  \begin{enumerate}[label=\alph*.]
	  
	 \item{\underline{{\bf Indemnification by Manufacturer}}. Manufacturer shall indemnify and hold harmless Buyer, its Master Distributors, affiliated and/or controlled companies, as well as each of their respective officers, directors, shareholders, agents, and employees, from and against all loss, liability, damages, claims for damages, settlements, judgments or executions, including costs, expenses and reasonable attorneys’ fees and costs (collectively, “Losses”) incurred by Buyer and/or such persons or entities as a result of any third party demands, actions, suits, prosecutions or other such claims arising on and after the date of this Agreement (“Third Party Claims”) based on: (i) any injury to or death of any person, or damage to property caused in any way by a Product provided by Manufacturer under this Agreement: (ii) any claims that a Product infringes any patent, copyright, trade mark right, trade secret, mask work right or other proprietary right of any third party, unless such claim is attributable to Manufacturer’s incorporation of formulations, specifications or materials provided by Buyer into the Products; or (iii) any alleged breach of Manufacturer’s representations and warranties contained herein.}
	 
	 \item{\underline{{\bf Indemnification by Buyer}}. Buyer shall indemnify and hold harmless Manufacturer, its subsidiaries, affiliated and/or controlled companies, as well as each of their respective officers, directors, agents, and employees, from and against all Losses incurred by Manufacturer and/or such persons or entities as a result of Third Party Claims based on: (i) any alleged breach of Buyer’s warranties contained herein, or (ii) any claims that a Product infringes any patent, copyright, trade mark right, trade secret, mask work right or other proprietary right of any third party to the extent such claim is attributable to Manufacturer’s incorporation of formulations, specifications or materials provided by Buyer into the Products. For purposes of this Section, all formulations, specifications or materials provided by Buyer into the Products shall be described in the Standards or on an attached memorandum signed by the Parties expressly referring to this Agreement.}
	 
	 \item{\underline{{\bf Indemnification Procedure}}. The party entitled to indemnification under this Section (the “Indemnified Party”) will provide the party obligated to provide indemnification under this Section (the “Indemnifying Party”) with prompt notice of any Third Party Claim for which its seeks indemnification under this Section, provided that the failure to do so will not excuse the Indemnifying Party of its obligations under this section except to the extent prejudiced by such failure or delay. The Indemnifying Party shall not be liable for any settlement effected without the Indemnified Party’s consent, which consent shall not be unreasonably withheld. The Parties shall cooperate in defending any Third Party Claim.}
	 
	  \end{enumerate}
	  
	  \item{\underline{{\bf Agency}}. Nothing contained herein shall be deemed to authorize or empower Manufacturer or its subsidiaries to act as an agent for Buyer or to conduct business in the name of Buyer.}
	  
	   \item{\underline{{\bf Entire Agreement}}. This Agreement, including its Exhibits and Purchase Orders issued under it, is the complete statement of the Parties’ agreement, and supersedes all previous and contemporaneous written and oral communication about its subject.}

	   \item{\underline{{\bf Compliance}}. Each Party will comply with all laws relating to the performance of this Agreement and represents and warrants that execution of this Agreement and performance of its obligations under this Agreement does not and will not breach any other agreement to which it is or will be a party, including but not limited to any agreements with its customers.}
	   
	   \item{\underline{{\bf Authority}}. The Parties represent that they have full capacity and authority to grant all rights and assume all obligations they have granted and assumed under this Agreement.}
	   
	     \item{\underline{{\bf Publicity of Agreement}}. The Parties agree that no press release or public announcement of this Agreement or concerning the activities contemplated herein shall be issued without the prior written consent of both Parties to the content of such release or public announcement, which consent shall not be unreasonably withheld.}
	     
	     \item{\underline{{\bf Further Assurances}}.The Parties agree to furnish upon request to each other such further information, to execute and deliver to each other such other documents, and to do such other acts and things, all as the other party may reasonably request for the purpose of carrying out the intent of this Agreement and the documents referred to in this Agreement.}
	     
	      \item{\underline{{\bf Attorney Fees}}. If any arbitration or legal proceeding is brought for the enforcement of this Agreement, or because of an alleged breach, default or misrepresentation in connection with any provision of this Agreement or other dispute concerning this Agreement, the successful or prevailing party shall be entitled to recover reasonable attorneys fees incurred in connection with such arbitration or legal proceeding. The term “prevailing party” shall mean the party that is entitled to recover its costs in the proceeding under applicable law, or the party designated as such by the court or the arbitrator.}
	     
	     \item{\underline{{\bf Exhibits}}. The exhibits attached hereto are an integral part of this Agreement and are specifically attached hereto and incorporated herein by this reference.}
	     \newline
	     
	     \end{enumerate}
	     
\noindent
	     
\noindent [Insert Signature Page]
\newline

\noindent EXHIBIT 1 - PURCHASE ORDER
[Insert Products to be purchased, including number, formulas, instructions, labels (if any), and location of delivery]
\newline

\noindent EXHIBIT 2 - PURCHASE PRICE
[Insert purchase price pursuant to purchase order (see Exhibit 1)]
\newline

\noindent EXHIBIT 3 - DELIVERY SCHEDULE
[Insert delivery date upon which Products are agreed to be delivered to Buyer]

%\end{enumerate}
\twocolumn

\end{document}